# A Security Suite for Wireless Body Area Networks

Raghav V. Sampangi[1], Saurabh Dey[2], Shalini R. Urs[3] and Srinivas Sampalli[1]

[1]Faculty of Computer Science, Dalhousie University, Halifax, Nova Scotia, Canada
`{raghav, srini}@cs.dal.ca`
[2]Analytics Quotient, Bangalore, India
`dey.saurabh84@gmail.com`
[3]International School of Information Management, University of Mysore, Mysore, India
`shalini@isim.ac.in`

## ABSTRACT

*Wireless Body Area Networks (WBANs) have gained a lot of research attention in recent years since they offer tremendous benefits for remote health monitoring and continuous, real-time patient care. However, as with any wireless communication, data security in WBANs is a challenging design issue. Since such networks consist of small sensors placed on the human body, they impose resource and computational restrictions, thereby making the use of sophisticated and advanced encryption algorithms infeasible. This calls for the design of algorithms with a robust key generation / management scheme, which are reasonably resource optimal. This paper presents a security suite for WBANs, comprised of IAMKeys, an independent and adaptive key management scheme for improving the security of WBANs, and KEMESIS, a key management scheme for security in inter-sensor communication. The novelty of these schemes lies in the use of a randomly generated key for encrypting each data frame that is generated independently at both the sender and the receiver, eliminating the need for any key exchange. The simplicity of the encryption scheme, combined with the adaptability in key management makes the schemes simple, yet secure. The proposed algorithms are validated by performance analysis.*

## 1. INTRODUCTION

Human health monitoring has greatly benefited from the recent advances in wireless communication and sensor technologies, which have enabled the use of sensors to assist in recording biometric data remotely. Sedentary lifestyles have increased the risks of potentially fatal medical conditions such as high blood pressure, cardiac diseases, diabetes, and the like, while hectic work schedules / absence of quality healthcare have contributed to increasing their risks. Given the unpredictable nature of worsening of any such conditions in a person, regular and continuous monitoring assumes high priority.

Wireless Body Area Networks (WBANs) [1] are a type of wireless sensor networks, where a group of sensors placed on the human body measure specific physiological parameters of a person and relay it to the monitoring medical center or hospital. This relay happens via the Internet or a cellular network, using personal digital assistants (PDAs) or cellular phones as intermediary devices. Thus, WBANs seem to be a promising solution for the problem of continuous health monitoring. However, with a patient's personal health data travelling in the open, typically in a wireless channel, to reach the intermediary device and then the monitoring station, securing this data becomes critical. This, coupled with the fact that medical decisions are made based on the data received, assumes significant focus in the research on WBANs.

To achieve security in any network, the messages to be transmitted are encrypted using specialized encryption schemes and a special encryption key, and decrypted at the receiver end. Many advanced encryption algorithms, which are used for securing wireless networks, however, cannot be used in WBANs, given that they have severe power constraints and resource constraints since they are small sensor devices residing on a person [1]. Thus, it becomes crucial to design algorithms that are simple in computation and resource utilization, yet achieve the desired security. At the heart of any encryption algorithm is the successful management of the special encryption key. The key generation scheme must also be computationally inexpensive yet secure.

This paper presents a security suite for WBANs, comprising of two dynamic key generation / management schemes for encryption of data in a WBAN, encompassing security in the part of the network from the central controller node of the WBAN to the receiver at the monitoring station, and between the sensors/actuators and the central controller node. The proposed schemes use the concept of random generation of keys for each data frame, keeping the mechanism of key generation and management simple, yet contributing to keep data secure.

The rest of this paper is organized as follows: Section 2 presents the related work. Section 3 introduces the security suite that is proposed, and describes the key management and encryption schemes in detail. Section 4 presents the performance analysis of the algorithms, while Section 5 provides a brief discussion on the schemes and their limitations. In Section 6, we conclude the paper and present a glimpse of the future work.

## 2. RELATED WORK

Security in WBAN is a critical issue given that the sensors transmit critical information related to the human body to the monitoring stations, and crucial decisions are made based on this data. If the data being communicated was intercepted by an adversary, and modified, it could prove fatal to the patient. Hence, this section of the paper discusses some of the work related to the area of security in WBANs.

Tan *et al* [2] present an identity based cryptography approach for security in WBAN. They identify the various security requirements in a WBAN, and present an adoption of the identity-based encryption (IBE) scheme, called IBE-Lite. In IBE, an arbitrary string is used to generate a public key, and the trusted third party derives the corresponding secret key separately. However, every time a new public/secret key pair is generated, the secret key must be stored in the trusted third party (Certificate Authority — CA), which then challenges the person who accesses the data. In IBE-Lite, a sensor generates a public key on the fly using an arbitrary string, but, the sensors cannot create secret keys. They, thus, use a trusted third party to ensure security of data in WBAN.

Moving away from the conventional key generation schemes and capitalizing on the inherent random and time variant nature of biometric data, Venkatasubramanian *et al* [3] present a scheme where the electrocardiogram (EKG) signals of a person are used to generate cryptographic keys for encrypting the communication between any pair of sensor nodes in a WBAN. In their work, sample values of EKG data are taken from a particular interval of the EKG signal, and using Fast Fourier Transform, coefficients are extracted. Using these coefficients, a feature vector is generated, which is used for generating the key. The derived key is then agreed upon by the sensors involved in the communication, and is used to encrypt the data in their communication. The main area of focus of this work is securing the inter-sensor communication within the WBAN. The sensors in the network agree upon a common key, generated based on the EKG reading of the patient. The keys generated using the presented scheme will be distinct for different people, since EKG data is different for different people; will be random, given the inherent randomness associated with biometric data; and, time variant. Mana *et al* [4, 5] also use electrocardiogram signals to generate secure keys between sensor nodes and the base station, thereby focusing on securing the end-to-end communication, as well as the communication among the nodes. Use of electrocardiogram signals in these works exploits the time-variant nature and the random behavior that it exhibits with varying physiological activities.

Raazi *et al* [6] propose a scheme that uses biometric measurements as symmetric keys (or private keys), since they are inherently random. In their scheme, they use the concept of key refreshment, where the personal server issues a key refreshment schedule periodically to all the sensor nodes in the WBAN. This refreshment schedule is used to change the key allotted to it for communication. Their scheme involves the use of three keys, namely—communication key, used for encryption in general communication in the network; administrative key, used to refresh the communication key; and, basic key, which is a key that is known only to the sensor node and the medical server that is used in the rare event of the administrative key being compromised. The initial key is pre-loaded to the nodes and periodically, the personal server chooses a value of a biometric to be used as a key for encryption by the nodes.

A notable point from the works discussed in this section is that all the encryption and key management schemes use keys that are exchanged between the transmitter and the receiver at one time or the other, or, use broadcasted keys in refreshment schedules. In such a scenario, if this key exchange communication was somehow intercepted, the attacker could easily get access to all the subsequent frames exchanged between the communicating entities. Thus, we can say that even though the system seems to be secure, it is still vulnerable.

The proposed security suite comprises of two key generation / management schemes that focus on introducing randomness to dynamically generate keys at both the sender and the receiver, independently. The schemes use the concepts of using multiple reference frames, the existing distinction in physiological data between people and periodic refreshes of such reference frames to achieve independent generation of keys at the sender and receiver. The security suite thus, forms the core of data security in a WBAN.

## 3. PROPOSED SECURITY SUITE

As explored earlier, WBANs seem to be a promising solution for the problem of continuous health monitoring. However, with a patient's personal health data travelling on a wireless channel to reach the destination, securing this data becomes critical. This, coupled with the fact that medical decisions are made based on the data received, assumes significant focus in the research on WBANs.

WBANs essentially consist of two sub-networks — the network of sensors on the human body, including all the individual sensors and culminating in the WBAN central controller (WCC), and the network between the WCC and the monitoring station. However, due to the resource restrictions in the sensors, security algorithms with the same level of sophistication cannot be implemented at both these sub-networks.

To overcome this, we propose a security suite, which uses a secure algorithm for the communication between the WBAN and the monitoring station, and a variant of the same algorithm for the network of sensors on the human body. Our work aims at reducing the resource utilization at both sub-networks, while maintaining the communication highly secure.

### 3.1. The Proposed Security Suite

Figure 1 illustrates the proposed security suite. We propose two algorithms namely, IAMKeys (Independent and Adaptive Management of Keys for Security in WBANs) [7] and KEMESIS (Key Management and Encryption for Securing Inter-Sensor Communication), to address the security needs of the different sub-networks.

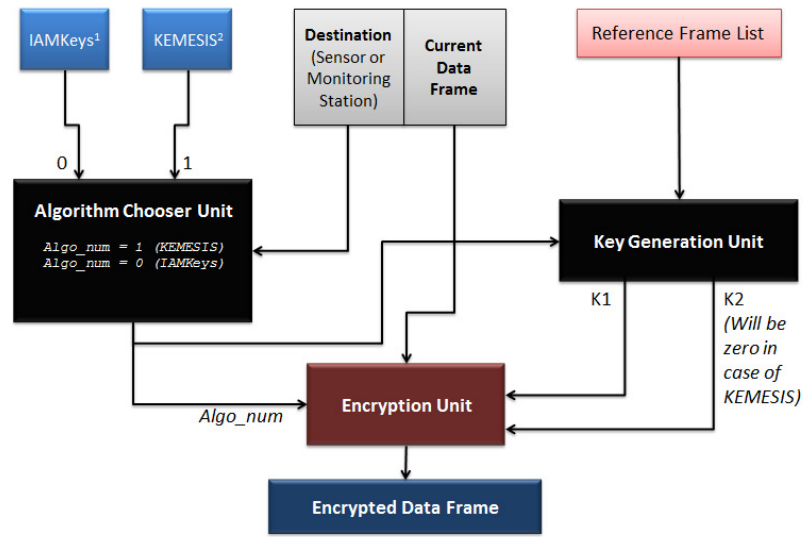

[1] *Independent and Adaptive Management of Keys for Secure Encryption in WBANs*
[2] *KEy Management and Encryption for Securing Inter-Sensor Communication*

Figure 1. The Proposed Security Suite

The WCC, being the interface between the two sub-networks, will have both the algorithms in its memory, while the sensors and the monitoring station will only have the necessary algorithms. Depending on whether the WCC is communicating with the sensors or the monitoring station, the "Algorithm Chooser Unit" will choose the appropriate key-generation algorithm, and enables generating the appropriate encryption keys. The key generation in both algorithms are dependent on a set of reference frames, and the details of key generation will be discussed in the following sections. The generated keys and the "Algo_num" (pointer to the chosen algorithm) are then used by the "Encryption Unit" to encrypt the data frame.

In case of IAMKeys, two encryption keys, K1 and K2, will be used, while in case of KEMESIS, only one key will be used, thereby further reducing computational overhead.

Having considered an overview of the proposed security suite, we now present the two key-generation and their corresponding encryption algorithms, which form the essence of the security suite.

## 3.2. IAMKeys: Independent and Adaptive Management of Keys for Security

To achieve security in any network, the messages to be transmitted are encrypted using specialized encryption schemes and a special encryption key, and decrypted at the receiver end. Most advanced encryption algorithms, which are used for securing wireless networks, cannot be used in WBANs. This is due to the severe power constraints and resource constraints resulting due to the small size of the sensor devices that reside on a person [1]. Thus, it becomes crucial to design algorithms that are lightweight — being simple in computation and resource utilization, yet achieving the desired security. At the heart of any encryption algorithm is the successful management of the special encryption key. The key generation scheme must also be computationally inexpensive yet secure.

Keys used in encryption play a crucial role in the security of data. If the key is compromised, it is as good as communicating in plain text. The significant part of the communication, thus, becomes the key exchange or the security association process. If an adversary eavesdrops on the packets associated with this process, then, the communication is compromised. However, if two entities were able to communicate without a security association process, it would resolve this issue. That is the focus of this scheme — to enable the sender and receiver to generate keys independently at either end, without the need to exchange keys.

IAMKeys is an independent key generation / management scheme for encryption of data in a WBAN, with focus on security in the part of the network from the WBAN central controller to the receiver at the monitoring station. This scheme focuses on generation of new keys for each iteration of encryption, which is accomplished at both the sender and the receiver, and further makes use of the slightly random nature of physiological parameters (such as heart rate) to make the keys random in nature. This, coupled with the use of an encryption scheme that combines the simplicity of stream ciphers and the complexity of block ciphers, makes the scheme efficient to secure the data communicated in a WBAN.

We consider the WBAN illustrated in Figure 2 for the purposes of this work. It includes sensors to measure the heart rate ($S_1$), the blood pressure ($S_2$) and the blood glucose ($S_3$) of a patient. The sensors send their recorded data (after basic filtering to remove the noise) to the WBAN Central Controller Node (WCC), which is marked as the sensor node $S_C$ in the figure. $S_C$ acts as the sink node for receiving the data on the body. The WCC then aggregates the readings in a data frame, along with a sequence number and a timestamp. Figure 3 illustrates the data frame structure. The WCC encrypts the data frame and forwards it to the medical monitoring station, via the intermediate personal device, which in our example is a cellular phone. We consider securing the portion of this WBAN system beginning at the WCC and culminating at the monitoring station.

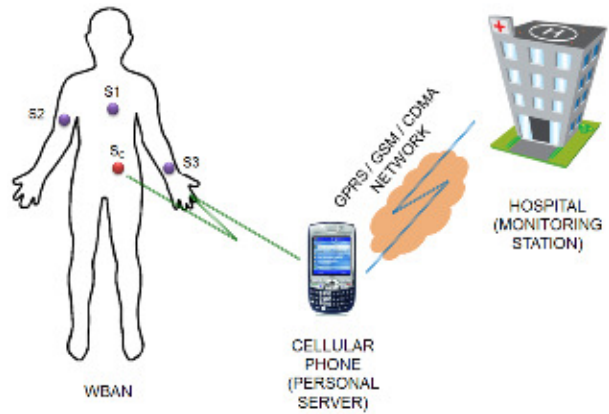

Figure 2. WBAN considered for the IAMKeys algorithm

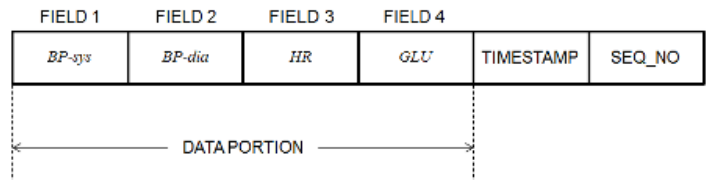

Figure 3. Data frame format considered in this work

Resource restriction in WBANs is one of the primary reasons that led us to develop a customized encryption and key management algorithm. Since the network is deployed on the human body, frequent maintenance activities, such as replacement of batteries, would prove highly inconvenient for patients. In case we manage to strike a balance between the computational expense and the algorithm, the next concern to be addressed is that of key exchange. In other proposals, any encryption key that is generated (or refreshed) is exchanged between the sender and the receiver. If an adversary were to eavesdrop on such a conversation, the entire communication between the sender and the receiver will be jeopardized and vulnerable to attacks.

An obvious solution to such an impasse is the use of symmetric cryptography. However, the key being constant in such schemes poses a threat to the system in question. Our work revolves around a scheme that:

- Removes the need for key exchange;
- Facilitates independent generation of keys at both sender and receiver;
- Ensures sender authentication;
- Simplifies the encryption process; and,
- Prioritizes freshness of data

To ensure the successful operation of such a scheme, we make the following assumptions:

- An administrator at the monitoring station (typically, a hospital) deploys the WBAN on the patient.
- At the time of installation, the administrator loads five "dummy" reference frames into both the WCC and the monitoring station data receiver.
- The receiver acknowledges every successful transmission.
- The sender receives the acknowledgement within the transmission of three subsequent frames.
- The study of the effects of radiation of the sensors on the human body has not been considered in this work.

Figure 4 illustrates the proposed scheme in brief. Each data frame is encrypted using a stream cipher based encryption scheme, with a key that is a pseudorandom number. A pseudorandom number generator (PRNG) generates the key using one of the randomly chosen data fields of one of the randomly chosen reference frames as the seed. The encrypted frame is then transmitted.

Upon reception, the receiver verifies the identity of the sender, and on successful authentication, generates the key independently to decrypt the data. Following this, the receiver sends an acknowledgement to the sender (WCC). Upon receiving the acknowledgement, the WCC updates the reference frame list by replacing the oldest reference frame (indicated by the sequence number) with the currently acknowledged data frame.

This scheme provides emphasis on data freshness. This is solely for the reason that with medical data being transmitted, retransmission of data acquired a while ago (lost due to collisions or packet losses), does not contribute to addressing the issue of continuous tracking. Therefore, instead of retransmission of lost packets, this scheme focuses on transmitting a frame with the latest data in case one of the previously transmitted data frames are not acknowledged by the receiver. Hence, as shown in figure 3, if frames 4 and 5 (FRM4 and FRM5) are lost due to some error in transmission, the sender will not retransmit these data frames, but, focus on transmitting the new data frame (FRM6), because latest data or current data is important.

This is a gist of the operation of the presented scheme. In the following sub-sections, we describe the constituents of the scheme in detail.

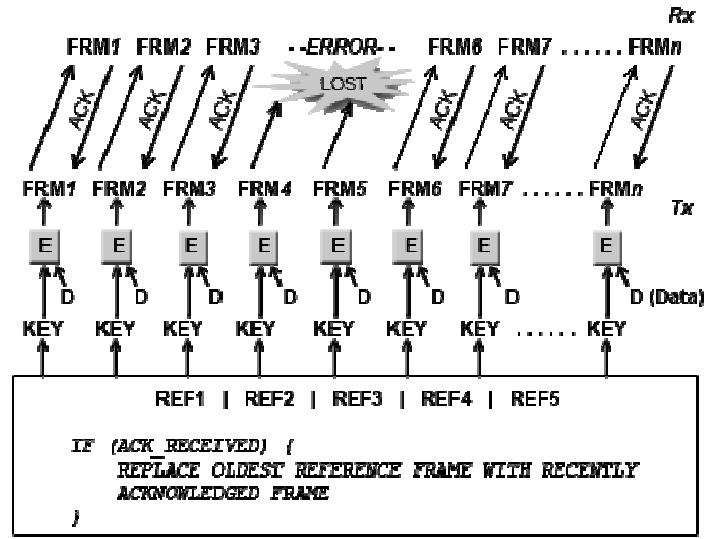

Figure 4.  Overview of the proposed scheme

### 3.2.1  Reference Frames

A crucial component of this scheme is the list of five reference frames that are a stored both at the sender and the receiver. During the WBAN deployment, the administrator loads these reference frames with identical dummy values at both the WCC and the monitoring station data receiver. When the receiver successfully receives the sent data frame, it sends an acknowledgement (ACK) to the WCC, along with the sequence number. Then, the receiver waits for one more reception, and updates its reference frame list. Upon reception of the ACK, the WCC proceeds to update its reference frame list. The reference frame list is updated as follows. The sender (or, receiver) checks for the oldest reference frame in its list. This is assumed to be the frame with the oldest sequence number. The system then replaces this reference frame with the acknowledged frame.

### 3.2.2  Sender Authentication and Tone

Each encrypted data frame that is transmitted is appended with the sequence number of the reference frame used for generating the key, the field number in this reference frame whose value is used as the seed, a "Tone" value, and a sender authentication code.

| ENC_DATA | SEQ_NO | REF_FRM_SEQ_NO | FIELD_NO | TONE | SENDER_AUTH |
|---|---|---|---|---|---|
| Encrypted current data | Current frame Sequence number | Reference frame Sequence number | Field in Reference frame chosen as seed | | Digital signature of the sender<br><br>Random integer between 1 and 5 |

Figure 5.  Structure of the transmitted frame

The transmitted frame is as illustrated in figure 5. The tone is a random number between 1 and 5, generated for each data frame. To generate the sender authentication code, the reference frame that was chosen to generate the encryption key is hashed the number of times indicated by the Tone value of that frame. This is analogous to generating a digital signature of the sender on the fly for each transmitted frame. The receiver repeats this process by retrieving the reference frame pointed by the reference-frame-sequence-number that was received as part of the transmitted message to authenticate the sender. The value of the tone has a minimum value of 1so as to ensure that the reference frame is hashed at least once, and has a maximum value of 5 to ensure that the hashing process does not become increasingly computationally expensive.

### 3.2.3  Key Generation and Management

After aggregating values for the data frame, the WCC proceeds to generate the keys for encrypting the current frame. It begins by randomly choosing one of the five reference frames, and one of the data fields of the chosen frame. The value of this field will be used as the seed for the PRNG, whose output will be the key, K1, which is logically inverted to generate key, K2. To avoid exchange of the keys, the WCC appends the sequence number of the reference frame and the field number in this frame as REF_FRM_SEQ_NO and FIELD_NO, respectively, in the transmitted frame.

The receiver, after authenticating the WCC, initiates the key generation process by first retrieving reference frame sequence number, followed by the field number. The receiver then retrieves the value of the particular field from its reference frame list, and uses this value (could be any biometric value in the reference data frame) as the initial seed for the PRNG to generate K1 and inverts it to obtain K2. The key generation operations are summarized in equations (1) and (2). These are the two keys that are used for the encryption and decryption processes, and are independently generated at both the sender and the receiver.

$$K1 = PRNG \text{ (SEED)} \tag{1}$$

$$K2 = INVERT \text{ (K1)} \tag{2}$$

where,
$PRNG$ () = Pseudorandom number generator; SEED = Value of the field pointed by FIELD_NO, in the reference frame pointed by REF_FRM_SEQ_NO; and, $INVERT$ () = Logical inversion operation

### 3.2.4  Data Encryption

Once the keys are generated, the WCC proceeds to encrypt the data. The encryption process is a combination of the concepts of block and stream ciphers, to ensure a simple encryption process that is also slightly complex. This proceeds as follows. At the time of data aggregation, the WCC assembles data as blocks of $k$ bits. This is to avoid the additional computation required to divide the aggregated frame into blocks of the specified size. The $k$ bit blocks of data and the keys, K1 and K2, are then divided into equal halves.

The encryption involves two rounds, one for each key. In the first round, the left half of the data is encrypted (XORed) with the right half of K1 to yield the right half of the intermediate frame, and the right half of the data is encrypted with the left half of K1 to yield the left half of the intermediate frame. The intermediate frame is the output of the first round of encryption. In the second round, the same logic as above is applied to the intermediate frame with K2. The left half of the intermediate frame is XORed with right half of K2 to yield right half of the encrypted frame, and the right half of the intermediate frame and left half of K2 are XORed to yield left half of the encrypted frame. Thus, after two rounds of encryption, the order of the data is preserved, but data is encrypted using a complex mechanism. Figure 6 summarizes the encryption process, and the operations are summarized in equations (3) and (4).

$$E1 = [ \text{ XOR (RHD, LHK1) } | \text{XOR (LHD, RHK1) } ] \tag{3}$$

$$E2 = [ \text{ XOR (RHE1, LHK2) } | \text{XOR (LHE1, RHK1)}] \tag{4}$$

where,

E1, E2 = Encrypted frames after round 1 and round 2, respectively; *XOR* () = The logical XOR function, that performs exclusive OR operation on its arguments; LHD, RHD = Left and Right halves of the Data frame; LHE1, RHE1 = Left and Right halves of the Intermediate encrypted frame (after round 1); and, LHK1, RHK1, LHK2, RHK2 = Left and Right halves of the keys, K1 and K2, respectively.

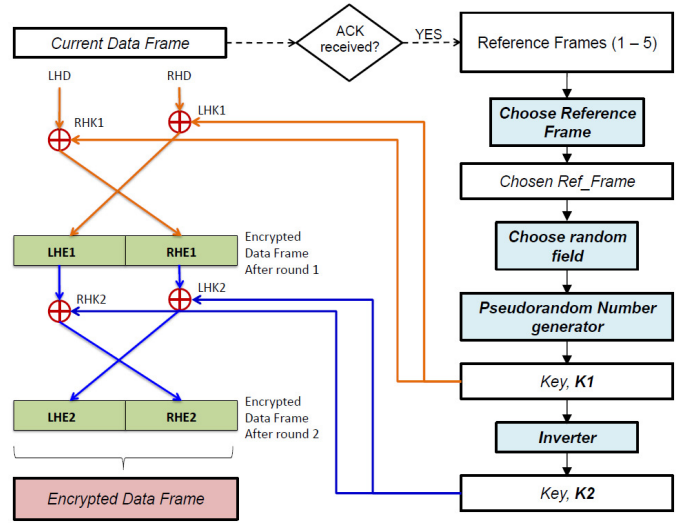

Figure 6.  Illustration of the Encryption process

### 3.2.5  Acknowledgements and Data Freshness

In applications that employ WBANs, such as healthcare, freshness (or recentness of data) is of utmost importance. Thus, even if a couple of frames are lost in transmission, and the WCC does not receive the corresponding acknowledgements, it continues to generate keys using the existing reference frames, and transmits the latest data frame encrypted using the keys generated. The concept of re-transmission of a lost data frame is not considered in such a scenario, with data freshness assuming priority. However, one needs to monitor the number of frames, which have not been acknowledged by the receiver. This is to avoid a case of lost connection between the WCC and the monitoring station, when several continuous frames are lost.

To address this, the WCC maintains a count of the number of frames, which have been transmitted since the transmission of a frame, say x1. If the acknowledgement is not received from the receiver within the transmission of 10 subsequent frames, the connection between the WCC and the monitoring station receiver is considered to be lost, and an alarm is raised. Meanwhile, if the receiver does not receive 10 consecutive frames, it considers the connection to be lost. In such a scenario, the administrator has to reset the connection.

### 3.3. KEMESIS: Key Management and Encryption for Securing Inter-Sensor Communication

The work presented in the previous section aims at securing the communication between the WCC and the receiver at the monitoring station. The possibility of employing such a scheme to protect the communication between the sensors and the WCC, referred to as the inter-sensor communication, would strengthen the communication further. In this section, we present a slightly modified version of the previously presented algorithm, taking into consideration the power and resource constraints of the sensors, to secure the communication between the sensors and the WCC. As in the previous scheme, this scheme removes the need for key exchange, enables independent generation of keys at both the sender and receiver, ensures seamless key refresh schedules at the sender and receiver, and simplifies the encryption process. To ensure the successful operation of such a scheme, we make the following assumptions:

- An administrator at the monitoring station deploys the WBAN on the patient;
- At the time of installation, the administrator loads n dummy reference frames into both the WCC and all the sensors of the WBAN, each having k fields

For the purpose of this work, we have considered the WBAN illustrated in figure 2, with the focus now being on the communication between sensors $S_1$, $S_2$, $S_3$ and the WCC, $S_C$. Figure 7 illustrates the structure of the $n$ dummy frames stored in the on-board memory of all the sensors and the WCC in the WBAN, and figure 8 illustrates the key management and encryption scheme used to secure the inter-sensor communication.

Let us assume that the sensors communicating in this instance are the Blood Pressure sensor ($S_2$) and the sink node (or, WCC, $S_C$). The scenario begins when $S_2$ wants to transmit its recently recorded data to $S_C$. It first randomly picks one of the $n$ dummy data frames, say $i$, and randomly picks one of the $k$ fields in this frame, say $j$. These are illustrated as actions of the $n$:1 multiplexers respectively, in figure 8.

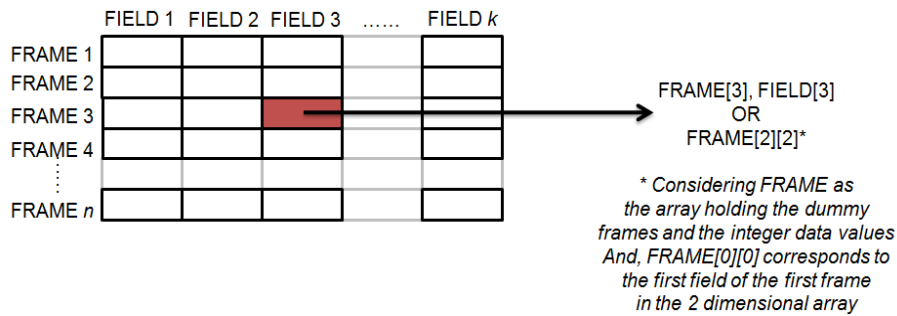

Figure 7.  Illustration of the *n* dummy frames stored as 2-Dimensional arrays in every sensor and the WCC of the WBAN

This field, i.e. FRAME[$i$][$j$], is chosen as the seed for the pseudorandom number generator (PRNG). The output of the PRNG is the key, K1. This key is logically inverted and the resulting key is K2. The actual encryption process is a simple XOR operation, with either of the keys, K1 or K2. In the following sub-sections, we discuss the crucial elements of this algorithm.

### 3.3.1 Dummy Reference Frames and Reference Frame Refresh Schedules

The core of this algorithm is the list of $n$ dummy reference frames (referred to as dummy frames because the data stored in these frames are randomly generated and bear no relation to the data being recorded by the sensors), stored in the on-board memory of all the WBAN sensors and the WCC. The value of $n$ can be suitably chosen, dependent on the application security needs and the computational power of each sensor. For example, a set of 256 frames, each with 256 bits, when divided into 16 fields of 16 bits each, lead to a possible $2^{16}$ values for each field, which increases the complexity and reduces the possible of hacking the value of the key generated based on this value. The administrator loads the values of these $n$ reference frames at the time of WBAN deployment in the memory of all the sensors and the WCC. If the values of the fields of these reference frames are refreshed periodically, the chances of cracking the algorithm reduce. Hence, the algorithm employs a key refreshment schedule.

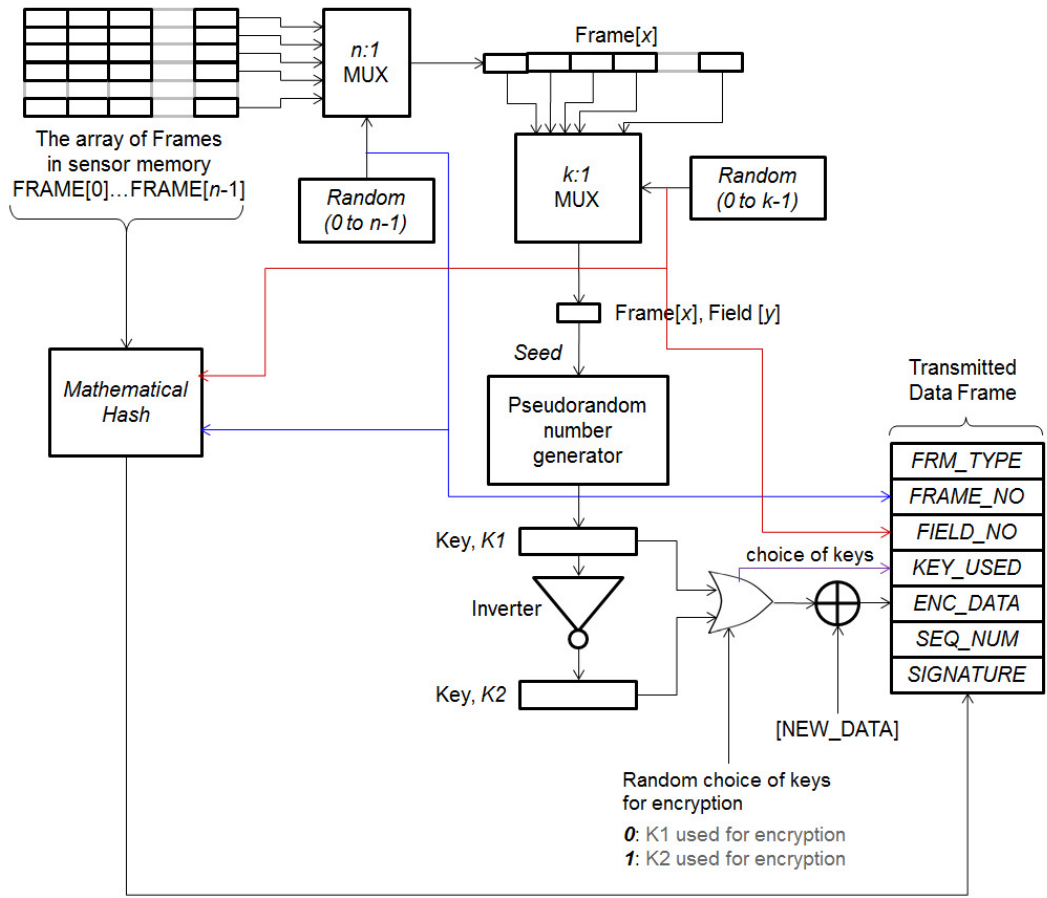

Figure 8. Illustration of the encryption and key management scheme employed to protect inter-sensor communication

The dummy reference frames are updated one field at a time, and in a random sequence. The transmitted frame indicates the type of frame, using the value set in FRM_TYPE field shown in figure 8. If the frame is a normal data frame (FRM_TYPE set to 0), the encrypted data (ENC_DATA) field holds the physiological data value being transmitted. If the frame is a control frame, the frame travels from either the sensors to the WCC or vice-versa. In this case, FRM_TYPE is set to 1, and the ENC_DATA field holds the value that will replace the value pointed by FIELD_NO, in the particular reference frame pointed by FRAME_NO. Thus, randomly, one of the fields of one of the randomly chosen reference frames is refreshed, thereby increasing the security of the algorithm.

### 3.3.2 Key generation and management

In this algorithm, the key generation process begins with the random choice of a reference frame from the set of reference frames. From this frame, one of the $k$ fields is randomly selected, and the value of this field becomes the seed for the PRNG. The output of the PRNG generated using this seed becomes the key, K1, and this key is logically inverted to obtain the key, K2, and either of these keys are randomly chosen for encrypting the data.

### 3.3.3 Sender Signature

To authenticate the sender, a simple one way mathematical hash function is used. The value of the field of the reference frame that was used as the seed for the PRNG is hashed by the sending sensor, and transmitted along with the encrypted data. On reception, the WCC first retrieves the data using the FRAME_NO and FIELD_NO values and computes the hash of the corresponding value in the reference frame. If the signatures match, the WCC continues with the decryption, and on successful decryption sends an acknowledgement. It has to be noted that the peripheral sensors will be in "sleep" state following transmission, and will be "woken up" by the acknowledgement signal.

In this algorithm, the acknowledgement frame helps in the sender and receiver synchronizing the data frame transmission, as well as synchronizing updates during control frame transmission. In case of control frames that are transmitted by the WCC, the sensors acknowledge successful reception, and in turn enable synchronization of updates.

### 3.3.4 Encryption

The encryption in this scheme is a simple XOR encryption scheme, employed taking into account the resource restrictions in the sensors. This is shown in figure 8. Once the keys, K1 and K2 are generated, either of them is randomly chosen and is XORed with the data that is to be encrypted, to yield the encrypted data. After encryption, the transmitter appends values to the KEY_USED field (0, if K1 is used for encryption and 1, if K2 is used for encryption); the FIELD_NO field with the field number in the reference frame pointed by the offset in the FRAME_NO field; and, the FRM_TYPE to indicate whether the frame being transmitted is a control frame or a data frame. This data frame is then transmitted.

In this section, we presented a novel algorithm that randomly generates and uses keys for encrypting each data frame in the inter-sensor communication in a WBAN. The algorithm randomly picks one of the fixed dummy reference frames and one of the random field values from it, and uses this value to generate a pseudorandom number sequence as the key. This key is inverted to generate the second key, and a choice of either key is made to find the final key used to encrypt the message. The encryption process is simple XOR encryption as in basic form of stream cipher based cryptosystems.

## 3.4. Summary: Connecting the dots

In the previous sections, we described the IAMKeys algorithm, and its variant, the KEMESIS algorithm, to secure the communication in a WBAN.

As illustrated in figure 1, the WCC chooses either IAMKeys or KEMESIS when communicating with the monitoring station or the other peripheral sensors of the WBAN, and encrypts the data using a combination of block and stream cipher concepts in IAMKeys, and a simple XOR function in KEMESIS. The receiver in both algorithms acknowledges receipt of the frame, thereby ensuring that the sender and receiver are synchronized to the same updates in the reference frames, which plays a central role in future key generation. In IAMKeys, the randomly generated tone value is used to add a digital signature to authenticate the sender, and since only the intended receiver can decode the frame and hence, send the acknowledgement frame, it achieves an indirect receiver authentication as well. In KEMESIS, however, since the sensors and the WCC reside on the human body, and to preserve resources, the need for a separate sender authentication process by using digital signatures is not considered in the current implementation.

In the next section, we present a performance analysis of the algorithms, in an attempt to justify the reduced power / resource utilization by the algorithms.

## 4. PERFORMANCE ANALYSIS

Preliminary analysis of proposed algorithms was done with a java program to emulate the desired operations. All the values / sensor readings were randomly generated using predefined mathematical random functions in java packages. In their work, Venkatasubramanian *et al* [3] present an analysis of randomness, time-variance and distinctiveness characteristics of the keys generated as performance analysis, based on which we present an analysis of our scheme.

*Randomness* is the unpredictable nature of the keys used. In the IAMKeys scheme, randomness of the keys generated by the PRNG is dependent on the seed, which is a function of three parameters—the chosen reference frame, the field in the reference frame, and the value of the field (physiological data). Since each of these values are randomly chosen for each data frame being encrypted, the randomness property of the seed, and hence the key, is preserved. We generated 100 frames of data, and noted the values of the offsets of randomly chosen reference frames and fields. Figure 9 illustrates the randomness of the reference frames and the field number values that were generated in IAMKeys. In case of KEMESIS, the randomness of the keys will be a function of the same three parameters, however, the value of the field is a randomly changing number, and not physiological data. Figure 10 illustrates the randomness observed when 16 reference frames, with 8 fields each were chosen to generate random keys. The randomness in KEMESIS is compounded by the fact that only one of the keys will be used for encryption.

*Distinctiveness* refers to how different the keys are when compared to different individuals. We observed from our simulation of the IAMKeys algorithm that keys are only identical for a person if the readings are exactly identical and the random reference frame or field values generated are the same, which is highly unlikely. In KEMESIS, the distinctiveness can be partly observed, given that only the WCC and the sensors of a given WBAN can have the same values in the reference frame list, since the key refreshes are random.

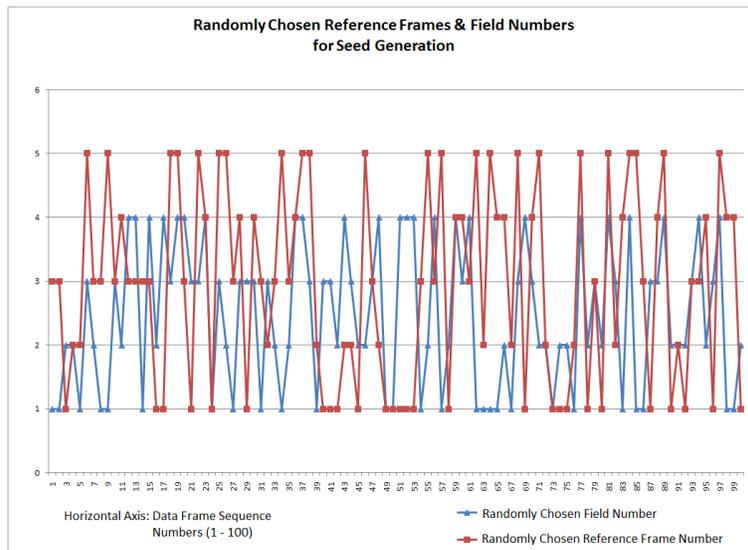

Figure 9. Simulation of reference frames and field numbers in IAMKeys

*Time-variance* of the keys implies that different instances of time should produce different keys. In a WBAN, not all sensor readings would vary drastically between intervals, and any drastic variation would indicate that the patient might be in distress. The time-variant nature of the keys in both algorithms is a function of the randomness property, and since the keys generated are random, we can say that they are time-variant as well.

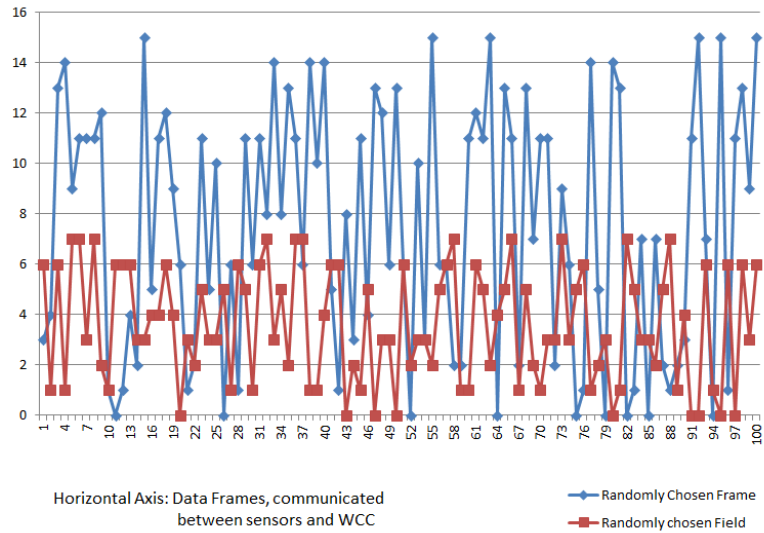

Figure 10. Simulation of reference frames and field numbers in KEMESIS

We analyze the complexity of the algorithm by determining the number of logical operations performed in the encryption and decryption of one data frame. However, for analysis purposes, we make the following assumptions:

- Data in each field of the frame is 8-bits.
- The PRNG is implemented using 8-bit Linear Feedback Shift Registers (LFSR).
- Reference frame and field numbers are randomly identified by numbers generated using 4-bit LFSRs.
- For addition operation, we consider the use of a *full adder* circuit for adding each bit, which employs 2 XOR gates, 2 AND gates and 1 OR gate. Hence, each bit addition will contribute to 5 logical operations [8].
- For sender authentication in IAMKeys, hash operation is performed on the sum of all data values in the reference frame. If $\alpha+1$ is the number of fields in the frame, then, there will be $\alpha$ extra additions during encryption/decryption.
- For sender authentication in KEMESIS, hash operation is performed only on the value in the field identified. Hence, $\alpha$ is not considered.
- We have assumed that reference frame list refresh and frame transmission do not count as logical operations, and thus, have not included these numbers in the calculations in table 2 or table 4.
- $\beta$ represents the tone value.
- $\gamma$ represents the presence or absence of a frame refresh following an acknowledgement frame reception.
- Hash operation is assumed to be one-bit circular left shift operation, contributing to one logical operation.

Table 1 highlights the various logical operations in key generation and encryption/decryption of one frame using IAMKeys. Table 2 presents the total number of logical operations, with calculations based on assumed values for $\alpha$, $\beta$ and $\gamma$ listed in table 1. The major contributing factor to the total logical operations is the sender authentication, which decides the value of $\beta$.

Table 1.  Number of logical operations in key generation and encryption / decryption of one frame in IAMKeys.

| Operation | Encryption [#] | Decryption [#] |
|---|---|---|
| *Random number generation*<br>Encryption: choice of reference frame, field in the reference frame, K1 generation, tone generation<br>Decryption: K1 generation | 4 | 1 |
| *Inversion*<br>Encryption & Decryption: Generation of K2 | 1 | 1 |
| *Exclusive OR (XOR)*<br>Encryption of each half of data in two rounds; and,<br>Decryption of each half of data in two rounds. | 8 x 4 | 8 x 4 |
| *Addition (or, increment)* [##]<br>Encryption: Sequence number, ACK monitor counter, addition of data fields in reference frame for hashing<br>Decryption: transmitted frame monitor counter, addition of data fields in reference frame for hashing | 8 x 5 x (2 + $\alpha$) | 8 x 5 x (1 + $\alpha$) |
| *Hash operation*<br>Encryption & Decryption: dependent on the tone value in the reference frame | $\beta$ | $\beta$ |
| *Reference frame refresh*<br>Encryption & Decryption: 0, if no ACK<br>1, if ACK received | $\gamma$ | $\gamma$ |
| *Frame transmission*<br>Encrypted frame and Acknowledgement frame transmission | 1 | 1 |
| **Total** | **118 + 40$\alpha$ + $\beta$ + $\gamma$** | **75 + 40$\alpha$ + $\beta$ + $\gamma$** |

# Addition and XOR operations are multiplied by 8, since they are bitwise operations.
## Since 5 logical operations are performed per bit addition, we multiply the factor by 5.

Table 2.  Number of logical operations in IAMKeys, with calculations based on assumed values of $\alpha$, $\beta$ and $\gamma$.

| Operating Scenarios | | Total logical operations | |
|---|---|---|---|
| | | Encryption | Decryption |
| Best case scenario | $\alpha = 3$; $\beta = 1$; $\gamma = 1$ | 240 | 197 |
| Average case scenario | $\alpha = 3$; $\beta = 3$; $\gamma = 1$ | 242 | 199 |
| Worst case scenario | $\alpha = 3$; $\beta = 5$; $\gamma = 1$ | 244 | 201 |

The performance for KEMESIS is very similar to that of IAMKeys, and is shown in table 3, while its operating scenarios are shown in table 4. The key points to note in this case are the absence of "tone", and block encryption.

Table 3. Number of logical operations in key generation and encryption / decryption of one frame in KEMESIS.

| Operation | Encryption [#] | Decryption [#] |
|---|---|---|
| *Random number generation*<br>Encryption: choice of reference frame, field in the reference frame, K1 generation<br>Decryption: K1 generation | 3 | 1 |
| *Inversion*<br>Encryption & Decryption: Generation of K2 | 1 | 1 |
| *Exclusive OR (XOR)*<br>Encryption and decryption of data | 8 x 1 | 8 x 1 |
| *Addition (or, increment)* [##]<br>Encryption: Sequence number, ACK monitor counter<br>Decryption: transmitted frame monitor counter | 8 x 5 x 2 | 8 x 5 x 1 |
| *Hash operation*<br>Encryption & Decryption: In KEMESIS, $\beta$ is always equal to 1, since the hash operation is performed only once per frame. | $\beta$ | $\beta$ |
| *Reference frame refresh*<br>Encryption & Decryption: 0, if no ACK 1, if ACK received | $\gamma$ | $\gamma$ |
| *Frame transmission*<br>Encrypted frame and Acknowledgement frame transmission | 1 | 1 |
| **Total** | **93 + $\beta$ + $\gamma$** | **51 + $\beta$ + $\gamma$** |

\# Addition and XOR operations are multiplied by 8, since they are bitwise operations.
\#\# Since 5 logical operations are performed per bit addition, we multiply the factor by 5.

Table 4. Number of logical operations in KEMESIS, with calculations based on assumed values of $\beta$ and $\gamma$. Note that $\alpha$ is not present in KEMESIS, since hashing is only based on the particular field value.

| Operating Scenarios | | Total logical operations | |
|---|---|---|---|
| | | Encryption | Decryption |
| Best case scenario | $\beta = 1; \gamma = 0$ | 93 | 51 |
| Worst case scenario | $\beta = 1; \gamma = 1$ | 94 | 52 |

# 5. DISCUSSION

With the primary objective of a WBAN being transmission of patient data with as little delay as possible, and such transmission being fool-proof, security and data freshness assume highest priority. In IAMKeys, we focus on data freshness with removing the need for retransmission of a lost frame. Instead of retransmission, the algorithm transmits the frame with the latest data values, hence, maintaining the data fresh. In case more than ten frames are lost or not acknowledged, the sender or the receiver will flag an error, and the administrator will need to check the communication link.

With security assuming high priority, optimizing resource utilization becomes a challenge. Tables 1 and 2 indicate that the resource utilization in IAMKeys varies mainly based on the value of β, which differs for each frame based on the randomly chosen reference frame. However, the actual implementation of hash operation and its complexity will vary the number of operations significantly. This gives the first limitation of this approach, where the complexity is not constant.

One of the primary objectives in our work is the optimization of resource utilization. An implementation specific complexity variation would then, decide the resources used by the algorithms, and in turn may lead to increasing the computational overhead for the WCC. Our work has focused on optimizing the computational overhead for the two component sub-networks individually, and an algorithm integrating the two, optimizing the computation for WCC is necessary. This would be implementation specific.

IAMKeys focuses on independently generating keys at both the sender and the receiver. The keys generated will be random for the encryption of each frame, due to the following reasons:

- Randomly, one reference frame out of the five in the list is chosen.
- A random data field is chosen in this reference frame.
- Even though the data may not be very random, the fact that such a data field is chosen randomly from a random reference frame, in addition to refreshing the list of reference frames, induces a sense of randomness in the keys generated.

KEMESIS is secure given its randomness in choosing the seed for encryption, and given that there is no need for an actual exchange of keys. Thus, similar to IAMKeys, the randomness of the keys is also a factor of the inherent randomness in the processes involved.

One of the important forms of attacks that we need to consider to analyze the security of such a system is man-in-the-middle (MITM) [10]. If an adversary were to listen to every conversation occurring in the system and modify the data as required, it would constitute MITM attack. Since randomness of the keys generated in the proposed schemes is a function of three parameters, the proposed algorithms have the property of dynamically changing encryption keys, which are not exchanged.

Another important attack that can be considered is the session hijacking [11], where the adversary initially becomes a part of the network, and gains control of the communication by assuming the role of either of the communicating entities. In IAMKeys, the decryption of each frame depends on the earlier successfully transmitted frames, which act as reference frames, while in KEMESIS, it depends on periodically updated reference frames. Further, since reference frames also authenticate the sender, it would be impossible for the adversary to pose as a sender with the absence of the refreshed reference frame list, thereby keeping data secure in IAMKeys. However, he may pose as the receiver, since the receiver is not acknowledged in the current implementation. This improves security of such a WBAN system and reduces the probability of an adversary taking control of the conversation.

Replay attack [12] is another important attack, involving the adversary using previously collected information from the communication between a legitimate WBAN and a monitoring station, to extract private information, or to hijack the session. Since both IAMKeys and KEMESIS employ the use of sequence numbers, replay attacks are avoided. It has to be noted that even if a couple of frames are lost in transmission, the sequence numbers are incremented, and with priority on freshness of data, the latest frame is transmitted. This further reduces the chances of a replay attack.

Further, since the fields of the reference frames (in KEMESIS) and the reference frames (in IAMKeys) are regularly refreshed, it can be said that even if an adversary were to capture some data frames, it would be difficult to decode those frames, and use them to decode any future frames. This ensures that future frames are secure, even if some of the packets in the communication were intercepted by the adversary. This contributes to increasing the security of the schemes, and hence, the proposed security suite.

One of the other limitations of the proposed schemes is relying on humans to ensure the randomness of the dummy (or initial) reference data frames. This can be avoided using an automated program to randomly assign dummy reference frames during the initial set up in the hospital. Another limitation is that it if the acknowledgement frame were lost, then, there would be no way for the monitoring station to know if the acknowledgement failed, and no way for the WCC to know that the monitoring station received the transmitted could, however, be avoided if there were two-way acknowledgement, where the transmitted frame would also contain the acknowledgement of the previously acknowledged frame by the receiver.

Though the proposed algorithms have some notable limitations, the chaotic nature that they impose on key generation keeps them stable and efficient, and less prone to attacks.

# 6. CONCLUDING REMARKS

In this paper, we have presented two key management schemes focusing on the independent generation of keys at the sender and the receiver. The need for the removal of exchange of keys in security mechanisms is fuelled by the fact that if the security association, or the initial phase of exchange of keys or exchange of numbers used to generate the keys, is compromised, the whole communication becomes vulnerable. In the proposed key generation / management schemes, we remove the need for key exchanges by using mechanisms that enable the sender and the receiver to generate the keys at their end. This ensures that the vulnerability of the communication at the security association phase does not apply when such a system is used.

Furthermore, since only the authorized sender / receiver will have access to the immediately previous medical data (which form the reference frames in case of IAMKeys) and to the periodically refreshed reference frames (in case of KEMESIS), it will be hard for any adversary to make any sense of the sniffed packets. Security increases with updates to the reference frames. The complexity is further improved by changing the encryption keys for each data frame. This ensures that the data frame remains secure under many given circumstances.

However, some notable limitations exist in the schemes, which will be the focus in our future work. It has to be noted that we have used the encryption schemes used, i.e. block based stream cipher encryption in IAMKeys and simple XOR encryption in KEMESIS, for the purposes of current implementation and analysis. Since the algorithms presented primarily emphasize on key generation / management, the use of other forms of encryption algorithms or a combination thereof will be a part of our future work. Our work has evolved around finding a balance between security and optimal resource utilization. With reduction in the total computational overhead, the proposed security suite is an initial attempt at optimally operating WBANs.

## Authors

**Raghav V. Sampangi**

Raghav Sampangi is pursuing his Ph.D. in the area of Security in RFID and Mobile Device Security at the Faculty of Computer Science, Dalhousie University, Halifax, Nova Scotia, Canada. His research interests include security and reliability in emerging wireless networks such as, RFID based networks, Wireless Body Area Networks, and Vehicular Ad-Hoc Networks. Specifically, he is interested in identifying security loopholes, and contributing to the research on security by addressing such issues. He has been an IEEE member since 2004.

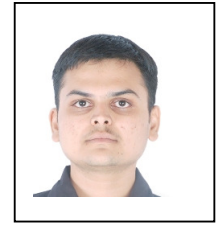

**Saurabh Dey**

Saurabh Dey is currently a Senior Programmer Analyst at Analytics Quotient, Bangalore, India. He has previously worked as a Lecturer at the Sikkim Manipal Institute of Technology, at Sikkim, India. His research interests include intelligence in wireless networks, wireless network security, and intelligent algorithms in text analytics. He is currently involved in research on Wireless Body Area Networks, as well as the development of applications on text and sentiment analysis.

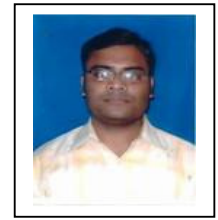

**Dr. Shalini R. Urs**

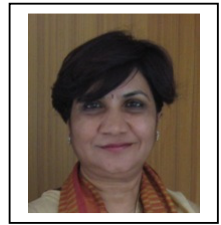

Dr. Shalini Urs is a Professor and the Executive Director at the International School of Information Management, University of Mysore, India. Her research ranges from the theoretical foundations of information science to technological aspects of information systems. Her areas of research include—Relevance and Information Retrieval, Information Systems, Ontology Development, and Social Media and Network Analysis, Data Analytics, and Information Security. She has been a faculty of the University of Mysore for the last 35 years and specializing in the areas of Information Systems and Management. She was a Fulbright scholar and visiting professor at the Department of Computer Science, Virginia Tech, USA during 2000- 2001. She has been researching into various aspects of Informatics, and published more than 100 papers in peer reviewed journals and prestigious academic conferences.

**Dr. Srinivas Sampalli**

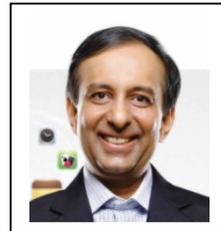

Dr. Srinivas Sampalli is a Professor and 3M National Teaching Fellow in the Faculty of Computer Science, Dalhousie University, Halifax, Nova Scotia, Canada. His research is mainly in the area of wireless security and applications. Specifically, he has investigated protocol vulnerabilities, security best practices, risk mitigation and analysis, design of intrusion detection and prevention systems, and applications of RFID systems and NFC-enabled smartphones. His projects have been sponsored by NSERC, Industry Canada and NRC. Dr. Sampalli has received many teaching awards at the Faculty, University, provincial and national levels, including a named teaching award and 3M National Teaching Fellowship, Canada's most prestigious teaching acknowledgement.